\documentstyle[11pt, epsfig]{article}
\newcommand{\blankline}{\vskip .3cm}
\newcommand{\f}{\begin{equation}}
\newcommand{\ff}{\end{equation}}
\newcommand{\bea}{\begin{eqnarray}}
\newcommand{\eea}{\end{eqnarray}}

\begin{document}
\centerline{\LARGE  Quantum geometry }
\centerline{\LARGE  with intrinsic local causality}
\blankline
\rm
\centerline{Fotini Markopoulou   
and Lee Smolin}
\blankline
\blankline
\blankline
 
\centerline{\it  Center for Gravitational Physics and 
Geometry}
\centerline{\it Department of Physics}
 
\centerline {\it The Pennsylvania State University}
\centerline{\it University Park, PA, USA 16802}
 \vfill
\centerline{December 2, 1997}
\vfill
\centerline{ABSTRACT}
 
The space of states and operators for a large class of background 
independent theories of quantum spacetime dynamics is defined.
The $SU(2)$  spin networks of quantum
general relativity are replaced by labelled compact 
two-dimensional surfaces. The space of states of  the 
theory is the direct sum of the spaces of  invariant 
tensors of a quantum group $G_q$ over all compact 
(finite genus) oriented 2-surfaces. 
The  dynamics is background 
independent and locally causal.
The dynamics constructs histories with discrete  features of
spacetime geometry such as causal structure and multifingered time. 
For $SU(2)$ the theory satisfies the
Bekenstein bound and the holographic hypothesis is recast
in this formalism.    

\blankline
email addresses: fotini@phys.psu.edu, smolin@phys.psu.edu 
\eject
\section{Introduction}

In this article we describe a new class of quantum geometries that 
is to be used in developing a theory of quantum gravity. 
These are natural extensions of the
spin network states that have been shown to comprise the
non-perturbative state space of quantum general 
relativity\cite{sn1,volume1}\footnote{
For recent reviews see \cite{carlo-review,future}.  
Spin networks were originally
introduced by Penrose \cite{roger-sn} as a model of quantum geometry.}.
In this formulation the labeled graphs on which spin 
networks are based are replaced by 
2-manifolds and the invariant tensors of a quantum group $G_q$
associated to them.   The motivations for this generalization 
comes partly from 
results of non-perturbative quantum gravity and string theory. 

This formulation has a kinematical part and a dynamical 
part.  The kinematical part, which is described in the
next four sections, generalizes the spin network
states in two ways.  The first is that the $SU(2)$ group of
the spin network states of quantum general relativity
is replaced by an
arbitrary quantum group $G_q$.    Within the framework
of non-perturbative, diffeomorphism invariant quantum
field theories, this is the natural
way to extend the degrees of the freedom of the theory
to include gauge fields\cite{spinnet-gauge} and supersymmetry
\cite{superloops}.  The quantum deformation is motivated from
physics by three considerations.  First, in quantum general relativity
the introduction of a cosmological constant $\Lambda$ requires
a quantum deformation of $SU(2)$ with $q=e^{2\pi /k+2}$ 
defined by\cite{linking,sethlee,srl}
\f
k= { 6\pi \over G^2 \Lambda } . 
\ff
Second, the truncation in the number of representations
with $q$ at a root of unity improves the formulation of
the dynamics by making the sums involved in the path
integral less divergent.  It also introduces new symmetries in the
theory which are not present in the classical case when
$q \rightarrow 1$.  These are analogous to the duality
symmetries of perturbative string theory.  
As we argue below, this  
may play a role in the interpretation of the theory.

The second sense in which our proposal extends the spin network
states of quantum general relativity is that the states are defined
intrinsically, without the use of a background manifold.  In
quantum general relativity the spin network states are 
diffeomorphism
classes of embeddings of graph in a fixed three-manifold
$\Sigma$ \cite{sn1,volume1}.  We go beyond this to a 
purely algebraic definition of
the state space which depends on no prior specification of a
manifold.  

One result of 
non-perturbative quantum gravity has been the 
discovery that geometrical quantities, including 
area\cite{spain,volume1},
volume\cite{spain,volume1,othervolume} and 
length\cite{thomas-length} have discrete spectra.  This is true 
before the introduction of 
 dynamics or matter couplings and signals that the combination
of diffeomorphism invariance and quantum theory requires that
quantum geometry be essentially discrete.  At the same
time, the application of these techniques to
the hamiltonian constraint of general relativity 
\cite{ham,roumen,qsdi} leads to a theory without
a good continuum limit\cite{trouble1,trouble2}.
Given this,
it seems more natural to construct the theory purely
from algebra and combinatorics and let continuum notions  
arise in the classical limit of the theory.   

We may note that the dualities of string theory suggest that 
one and the same physical situation may sometimes 
be described in two
different ways, which differ in the topology and manifold structures
of the underlying manifolds\cite{Sduality}.  
Other results show that in string theory
there are continuous phase transitions whose semiclassical 
description involves abrupt changes of the topology of the
underlying manifold\cite{mirror}.  These 
suggest that the fundamental,
non-perturbative, description should not be based on 
fixed topological manifolds.  

But without background manifolds the theory cannot be 
formulated in terms of the embeddings of surfaces or membranes.
The alternative is to construct the 
states and operators that are to represent quantum
geometry algebraically, using only
combinatorics and representation theory.  This is the main goal
of this paper.  In
\cite{pqtubes} and \cite{stringsfrom} results are
presented consistent with the hypothesis \cite{future} 
that the resulting extension of the spin networks 
formalism may serve as
framework for non-perturbative string theory.\footnote{
In fact the basic idea of the 
present formulation is rather like the
idea behind the transition from quantum field theory to
perturbative string theory.  Just as Feynman diagrams are
replaced by string worldsheets, the present generalization 
of quantum general relativity extends spin network states
to 2-dimensional surfaces  and the states of
field theories defined on the surfaces.}

A theory formulated without reference to any background
manifold still requires  dynamics and that dynamics 
should have built
into it some notion of local causality.  Below, in section 7,
we show that this can be achieved by an extension of
a formulation of spin network dynamics proposed earlier
by one of us\cite{fotini1}.  
The dynamics is based on discrete histories
$\cal M$, which are combinatorial
structure which have two properties shared by classical
spacetime:

\begin{enumerate}

\item{}Each history $\cal M$ 
contains a finite set $\cal E$ of elements that may be
called ``events".  This set of events is a partially ordered
set. We thus have  the finite element analogue of
the points of a Lorentzian
spacetime. 

\item{}Each history $\cal M$ contains a large number of
connected sets of causally unrelated events, which may be
called ``quantum spacelike surfaces".  Each spacelike surface
is also a quantum state.  Thus, the theory has a 
discrete analogue of the many-fingered time of general
relativity, which means that a discrete analogue of
spacetime diffeomorphism invariance is built in.

\end{enumerate}

Each history is then given an amplitude which is a product of
factors each associated to a local transition in the quantum
geometry. Causality and locality impose restrictions on the
choice of these amplitudes which are discussed below.
The issue of the choice of dynamics and the related
question of the
continuum limit is discussed in \cite{fmls1, stringsfrom}.

In the next three sections we introduce the space of states that
we propose extends spin networks and describe useful decompositions
of them which are based on 3- and 4-punctured spheres.
Section 5 introduces an algebra of operators that act on the
states and the interpretations of some of them, which yields a
picture of quantum geometry, is the subject of section 6.  
The dynamics of the theory is described in section 7,
while section 8 discusses coarse-grained observables and entropy
and their relationship to the holographic hypothesis and Bekenstein
bound.  
The conclusion is
largely devoted to describing ongoing work that will be reported
in other papers.

\section{The space of states}

The space of states that we investigate here is both the 
 extension of $SU(2)$ spin networks to a  
 quantum group $G_q$ and of the spin network states of canonical 
quantum gravity to the non-embedded case.

Given a  quantum group $G_q$ and a compact oriented 2-surface
${\cal S}$, let ${\cal V}^{{\cal S}}_{G_q}$ be the space of $G_q$ 
invariant
tensors on $\cal S$.\footnote{
Equivalently this is the space of
conformal blocks of the {\small WZW} theory corresponding to level $k$
on $\cal S$ \cite{ms,louis2d3d,other2d3d}
or the space of states
of $G_q$ Chern-Simons theory on $\cal S$, seen as a spatial
slice of some 3-manifold\cite{witten-cs}.}
We then define the space of states
${\cal H}_{G_q}$ of $G_q$ quantum gravity to be
\f
{\cal H}_{G_q} = \bigoplus_{\cal S} {\cal V}^{{\cal S}}_{G_q}
\label{states}
\ff
where the sum is over all compact 2-surfaces of finite genus.
Each
${\cal V}^{{\cal S}}_{G_q}$ is finite dimensional when $q$ is at a 
root of unity. ${\cal H}_{G_q}$ is 
equipped with the
natural inner product (see (\ref{ip1}) below) and is
a Hilbert space.

The sense in which these states may be considered to constitute 
an extension of the spin network states of quantum general relativity
will be discussed shortly, but we note that this is not a new notion.
It is known that the quantum deformation
of spin networks 
requires that their edges be enlarged to ribbons or 
tubes\cite{witten-cs,louis2d3d,other2d3d}.
This is to allow dependence of the states on twistings of the edges,
necessary for the $q$-deformed 
case\cite{witten-cs,louis2d3d,lou-sn,ribbons}.  
In the next sections we 
investigate properties of these states that
are important for their physical interpretation.

\section{Trinion decomposition: basis states}

\begin{figure}
\centerline{\mbox{\epsfig{file=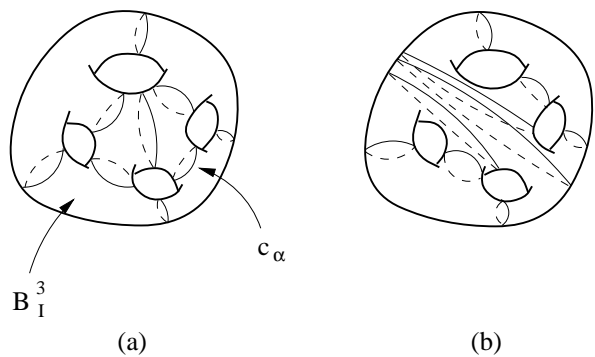}}}
\caption{(a) A genus 4 surface cut into six trinions $B^3_I$ by circles
$c_\alpha$. (b) The same surface in a different trinion decomposition. } 
\label{trinions}
\end{figure}

We begin by reviewing some of the properties of the
state spaces ${\cal V}^{\cal S}_{G_q}$ that we will 
need to discuss their
role in representing the states of quantum 
gravity\footnote{Complete characterizations of
 ${\cal V}^{\cal S}_{G_q}$ may be found in \cite{ms,louis2d3d}.}.
For the purposes of describing the states and operators
on ${\cal H}_{G_q}$ it will be very useful to understand
the behavior of the states in
${\cal V}^{\cal S}_{G_q}$ under decompositions of the surface
$\cal S$ into a union of punctured spheres.  
We begin by discussing the decomposition of a 
genus $g$ surface $\cal S$ into 
3-punctured spheres, or trinions.  Given a surface $\cal S$ we may
choose a maximal set of non-intersecting elements of $\pi^1 [{\cal S}]$,
which we shall call circles, $c_\alpha$.  
Cutting $\cal S$ along the circles $c_\alpha$ decomposes it into
a set of $N$  trinions, $B^3_I$, $I=1,...,N$. 
The trinions are joined on their punctures
so that each circle $c_\alpha$ corresponds to the punctures on two
trinions. (See Figure \ref{trinions}a.) This may be done in several
different ways.  (See Figure \ref{trinions}b.)

A trinion decomposition will be called {\it non-degenerate} if no
two trinions meet at more than two circles
(see Figure 2).  

\begin{figure}
\centerline{\mbox{\epsfig{file=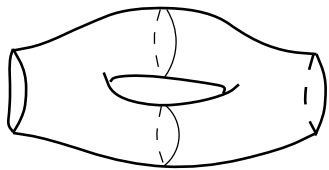}}}
\caption{This trinion decomposition is degenerate because the two trinions 
have two circles in common. } 
\label{degenerate}
\end{figure}

Associated to each trinion decomposition of $\cal S$ is a 
class of bases of
${\cal V}_{G_q}^{\cal S}$, which is constructed as follows.  
$G_q$ has a list of irreducible
representations, which we will label by $j_\alpha$.
(For the $q$ taken at a root of unity, which we will assume,  
this is a finite list.)  Each of the three
circles of a trinion $B^3_I$ may be labeled by a representation
$j_\alpha$, $\alpha=1,2,3$.  For each choice of the representations
$j_\alpha$ there is a linear space ${\cal V}_{j_1j_2 j_3}^I$
of intertwiners $\mu_I$. The intertwiners are the maps
\f
\mu_I:j_1\otimes j_2\otimes j_3 \longrightarrow {\bf 1}.
\ff
A choice of a set of $j_\alpha$  on the punctures of a trinion
is called consistent if
the corresponding ${\cal V}^I_{j_1j_2 j_3}$ has strictly positive 
dimension.

The space of states ${\cal V}_{G_q}^{\cal S}$ (subset of 
${\cal H}_{G_q}$) 
associated with the surface
$\cal S$ is constructed by taking direct products of all the constituent
spaces ${\cal V}^I_{j_1j_2 j_3}$ and summing over the representations,
\f
 {\cal V}_{G_q}^{\cal S}= \sum_{j_\alpha}\bigotimes_I 
{\cal V}^I_{\{ j\}_I}
\label{decom}
\ff
where $I$ labels an arbitrary trinion $B_I^3$ in
$\cal S$ with labels $\{ j\}_I$. 

A generic state in  ${\cal V}^{\cal S}_{G_q}$ will be
denoted $|{\cal S}, \Psi \rangle $.  
A basis in ${\cal V}^{\cal S}_{G_q}$ is
then constructed as follows.  We choose
an orthogonal basis of intertwiners in the
space ${\cal V}^I_{\{ j\}_I}$ of each of the trinions, denoted
$\mu_I^\rho$.  A basis of states in $ {\cal V}_{G_q}^{\cal S}$ is then
given by a choice of $j_\alpha$ on each circle $c_\alpha$ in ${\cal S}$
and a choice of a basis element $\mu_I^\rho$ on each trinion.
These basis states are denoted 
$|{\cal S},j_\alpha , \mu_I^\rho \rangle$.

Given a trinion decomposition of every finite genus 
2-surface $\cal S$, the
states  $|{\cal S},j_\alpha , \mu_I^\rho \rangle$ provide an
orthonormal basis for the state space ${\cal H}_{G_q}$.
The inner product on ${\cal H}_{G_q}$ is given by 
\f
\langle{\cal S},j_\alpha , \mu_I^\rho 
|{\cal S}^\prime,j_\alpha^\prime , \nu_I^\tau \rangle
=\delta_{{\cal S}{\cal S}^\prime}\prod_\alpha 
\delta_{j_\alpha j_\alpha^\prime}\prod_I 
\langle\mu_I^\rho|\nu_I^\tau \rangle_I
\label{ip1}
\ff
where the same trinion decomposition is assumed for the
two states when ${\cal S}\cong{\cal S}^\prime$ and 
$\langle\mu_I^\rho|\nu_I^\tau\rangle_I$
is the inner product in the space of intertwiners 
${\cal V}^I$ on the $I$-th trinion.

\begin{figure}
\centerline{\mbox{\epsfig{file=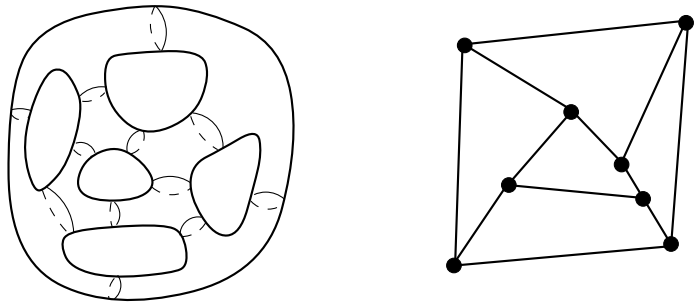}}}
\caption{A trinion decomposition of a genus 5 surface reduced
to a spin network graph.} 
\label{tubenet}
\end{figure}

Note that, given a particular trinion decomposition of $\cal S$, the
states in the basis $|{\cal S},j_\alpha , \mu_I^\rho \rangle$
may be thought of as generalized combinatorial 
trivalent spin networks. (See Figure \ref{tubenet}
but note that for $q\neq 1$
these are quantum spin networks\cite{lou-sn}.).
The edges $e_\alpha$ of the corresponding graph $\Gamma$ are  
labeled with the same 
representations $j_\alpha$ as the corresponding 
circles $c_\alpha$, while the trivalent nodes $v_I$
associated to the trinions are 
labeled by the intertwiners $\mu_I$.  Because of this association
we sometimes call the basis states 
$|{\cal S},j_\alpha , \mu_I^\rho \rangle$ 
{\it tubular spin networks}. 

The assignment of a graph $\Gamma$ to the surface $\cal S$ 
 depends on the choice of the trinion decomposition
and the same is thus true of 
the basis $|{\cal S},j_\alpha,\mu_I^\rho \rangle$.
If we choose a different
trinion decomposition of $\cal S$, based on a different
maximal set of non-intersecting elements of $\pi^1 [{\cal S}]$,
we have a different basis for ${\cal V}_{{\cal S}}^{G_q}$.
The recoupling identities of the representation
theory of $G_q$ \cite{lou-sn} then provide 
the change of basis formulas.
Alternatively, they may be computed using the modular
transformations of the corresponding rational conformal field theory
as in \cite{ms,louis2d3d}.

We may note that when $q \rightarrow 1$ the spaces 
${\cal V}_{G}^{\cal S}$ become infinite dimensional as there
are an infinite number of representations $j_\alpha$.  Then,
the Moore-Seiberg operators are no longer well defined
unitary operators.   Thus, in the limit
$q \rightarrow 1$ the states in ${\cal H}_{G_q}$ are
the usual combinatorial spin network states of $SU(2)$.
 
\section{Decomposition in 4-punctured spheres}

Just as the trinion decomposition is related to 
an extension of trivalent
spin networks, we can associate an extension of 4-valent
spin networks to the bases of states in 
${\cal V}^{\cal S}$ that come from decomposing $\cal S$
into 4-punctured spheres (from now on we 
drop the suffix $G_q$ of
${\cal V}^{\cal S}_{G_q}$).  To accomplish this we pick a
(non-maximal) set of non-intersecting circles $c_\alpha$ on $\cal S$
that decompose it into 4-punctured spheres $B^4_I$.  
As before we can label these circles with representations
$j_\alpha$.  

It will also be useful to work with general $n$-punctured spheres.
In general, a 
2-sphere with $n$ punctures, denoted 
$B^n_I$, is
labelled by representations $j_1,...,j_n$ of the group $G_q$.
Given $B^n_I$ and the labels $j_1,...,j_n$, there
is a linear space of
intertwiners, ${\cal V}^I_{j_1,...,j_n}$, consisting of the invariant
maps
\f
\mu_I:j_1\otimes ... \otimes j_n\longrightarrow {\bf 1}.
\ff
As in the 3-punctured case, the 
dimension of ${\cal V}^I_{j_1,...,j_n}$ is required to
be non-zero otherwise the choice of $j_1,...,j_n$ is
inconsistent and not allowed.
Now, given any decomposition of $\cal S$ into $n$-punctured 
spheres along a set of circles $c_\alpha$,  we have 
representation of the states in ${\cal V}^{\cal S}$
in terms of triples $|{\cal S},j_\alpha , \mu_I\rangle $.  The formulas
(\ref{decom}) and (\ref{ip1}) still hold.  

Returning to the decompositions in terms of 
4-punctured spheres, we may note that such decompositions
may be made, at least locally, in a surface $\cal S$ by grouping
the trinions in some trinion decomposition into pairs.
Finally, as in the case of trinions, we call a 
decomposition of a surface $\cal S$
into 4-punctured spheres {\it non-degenerate} if no two
4-punctured spheres share more than one puncture.

\subsection{The tubular 4-simplex}

\begin{figure}
\centerline{\mbox{\epsfig{file=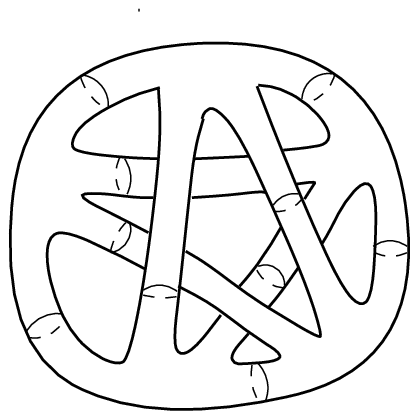}}}
\caption{The tubular 4-simplex ${\cal P}$, a genus 6 surface decomposed 
to 5  4-punctured spheres. } 
\label{4simplex}
\end{figure}

In fact, a genus $g$ surface always has a non-degenerate decomposition 
into 4-punctured spheres for $g \geq6$.
It is easy to see that the smallest number of $4$-punctured
spheres that can fit together non-degenerately is $5$.  These
make up a genus $6$ surface which may be thought of as
a tubular generalization of the 4-simplex (see \cite{fotini1}), 
as every
4-punctured sphere $B_I^4$, $I=1,...,5$ is connected to every
other one once.  (See Figure \ref{4simplex}).  This surface plays a special
role in the dynamics. We shall call it  $\cal P$ and
refer to as
the {\it generating surface}.  
Together with  ten fixed circles $c_{IJ}$ connecting $B^4_I$
and $B^4_J$
that decompose it into five such 4-punctured spheres
$\cal P$  will be called the
{\it tubular 4-simplex}.  
Its $q \rightarrow 1$ limit is a 4-valent
graph with 4-valent nodes $v_I$ for each
4-punctured sphere $B^4_I$ of ${\cal S}$ and an edge $e_{IJ}$ for
for each circle $c_{IJ}$.
 Labeling the circles $c_{IJ}$ by
representations $j_\alpha$ and the $B^4_I$ by a basis
$\mu_I^\rho$ in the corresponding spaces of intertwiners
${\cal V}^{B^4_I}_{\{j_\alpha\}}$ we have a basis of states
$|{\cal P} , j_\alpha, \mu_I^\rho \rangle $.  Each of these
is a coloring of the 4-simplex.  

\subsection{Tubular evolution moves}

Consider a non-degenerate decomposition of a surface 
$\cal S$ into $n$ 4-punctured spheres,
\f 
{\cal S}=\bigodot^n_{I=1} B_I^4, 
\ff
where $\odot$ denotes the gluing of a pair of punctures
with the same labels. 
Given ${\cal S}$, there is a set of 
local moves each of which yields another surface ${\cal S}^\prime$
expressed as a non-degenerate composition of 4-punctured
spheres ${\cal S}^\prime=\bigodot^m_{I=1} B_I^4$ where
in general $m \neq n$.\footnote{
These moves are a generalization of the 
Pachner moves from combinatorial topology 
\cite{pachner} that played
an important role in the evolution of spin networks in \cite{fotini1}.}

To define these moves let us now put forward some notation.
An {\it elementary local region}, $L$, is a set of $n \leq 4$
4-punctured $2$-spheres 
$B_I^4$,
\f
L=\bigodot^n_{I=1} B_I\qquad n\leq 4,
\label{eq:L}
\ff
each pair of which is connected by exactly one tube.  $L$, therefore, 
is a 2-surface with 4 or 6 punctures, their number given by
\f
{\rm number\ of\ free\ punctures}=
4n-n(n-1).
\ff\label{eq:L2}
$n$ has to be at most 4 for $L$ to have any free punctures. 
For $n=1,2$ the genus (not counting punctures) 
of $L$ is 0, for $n=3$ it is 1 and for 
$n=4$ it is 3.  
 
\begin{figure}
\centerline{\mbox{\epsfig{file=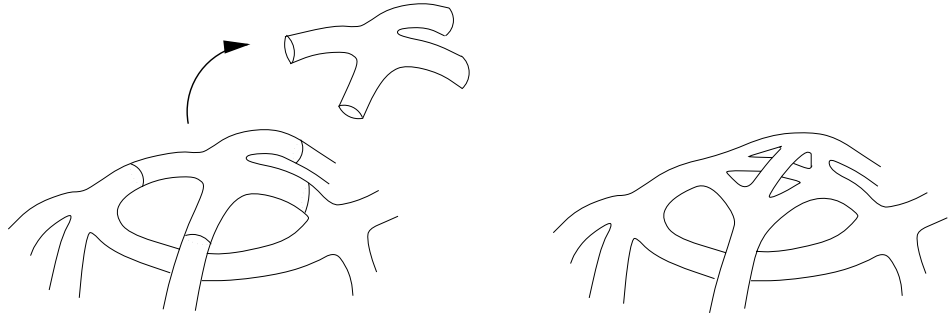}}}
\caption{An elementary substitution move.} 
\label{move}
\end{figure}

Given these definitions, 
a local move is the following.   Given 
a decomposition ${\cal S}=\bigodot^n_{I=1} B_I^4$, remove an 
elementary local region
$L$ in it and replace it with a new one $L'$ that has the same
number of punctures and same labels on the punctures. 
(See Fig. \ref{move}).

\begin{figure}
\centerline{\mbox{\epsfig{file=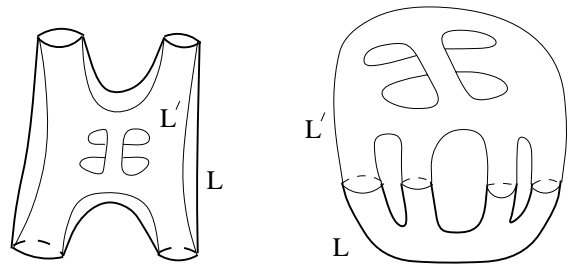}}}
\caption{Left:  The substitution move seen as a three manifold
that defines a cobordism from $L$ to $L^\prime$.
Right: Joining the $L$ and $L^\prime$ together makes
a generating surface $\cal P$.  } 
\label{move2}
\end{figure}

The topology of the new local region is determined by requiring that
$L$ and $L'$ can be composed along their common
punctures to form the generating surface, $\cal P$ 
(see Figure(\ref{move2})),
\f
L'\odot L={\cal P}_{L'\odot L}.
\label{eq:LL'}
\ff

\begin{figure}
\centerline{\mbox{\epsfig{file=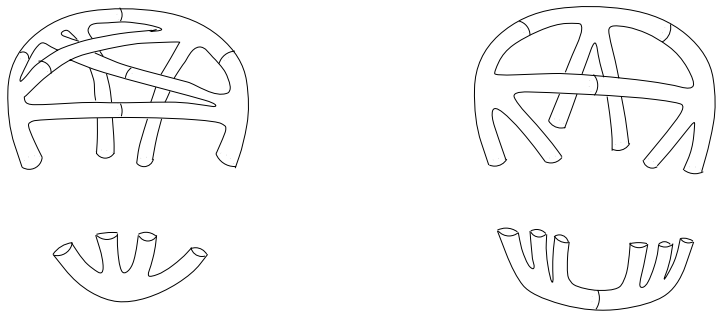}}}
\caption{The four elementary substitution 
moves, $1\leftrightarrow 4$ and $2 \leftrightarrow 3$.} 
\label{move3}
\end{figure}

Namely, $L'$ is the complement of $L$ in ${\cal P}_{L'\odot L}$.
\footnote{
In terms of the vector space representations of $L$ and $L'$,
equation (\ref{eq:LL'}) is the tensor product of the vector space
of $L$ with the dual vector space of $L'$, 
${\cal V}^L\otimes \left({\cal V}^{L'}\right)^D$.} 
We call such a substitution a {\it tubular evolution move.} 
(See Figs.\ \ref{move}, \ref{move2} and \ref{move3}). The
result is a new 2-manifold ${\cal S}^\prime$ which has
a decomposition into 4-punctured spheres that fall into
two sets, those in ${\cal S}-L$ and those in $L^\prime$.  
 
It is clear that if the original decomposition of $\cal S$ into
4-punctured spheres is non-degenerate then so is the
new one. No two $4$-punctured spheres in
${\cal S}-L$ share more than one connection because the original
decomposition is non-degenerate.  The same is true for
the $4$-punctured spheres in $L^\prime$ because every 
elementary local region is non-degenerate.   The
non-degeneracy of $L^\prime$ implies that there can be
at most one connection between any sphere in $L^\prime$
and one in ${\cal S}-L$.
\footnote{
This is an extension of the basic
fact that the Pachner moves applied on a $PL$ triangulation takes
preserves non-degeneracy of triangulations, i.e. when 
no two tetrahedra share more than one face.}

There are four kinds of tubular evolution moves, depending
on the number of $4$-punctured spheres in the old and new
elementary regions $L$ and $L^\prime$. As in the case of  the
Pachner moves used in \cite{fotini1}, these are denoted the
$1 \rightarrow 4$, $4 \rightarrow 1$, $2 \rightarrow 3$ and
$3 \rightarrow 2$ moves.  In terms of the corresponding surfaces
one can see from Figure \ref{move3} that these result in a change of
genus by $+3$, $-3$, $+1$ and $-1$ respectively.   This means that
starting with the tubular 4-simplex $\cal P$ which has genus
$6$, one can make $r$ successive $2 \rightarrow 3$ moves to reach
a surface of any genus $g=6+r$.  Therefore, each surface
with genus $g \geq 6$ has a non-degenerate decomposition 
into 4-punctured spheres.

Note also that
if $B^4$ is a 4-punctured sphere with
labels $j_1,j_2,j_3,j_4$ it may be decomposed along a
circle $c_1$ into two trinions $B^3_1$ and $B^3_2$.  If we
call the label on $c_1$ by $l$ we have
\f
{\cal V}_{j_1 j_2 j_3 j_4}^{B^4} = \sum_l 
{\cal V}_{ j_1j_2l}^{B^3_1} \otimes
{\cal V}_{l j_3j_4}^{B^3_2}.
\ff
 Thus, we see that
there are many trinion decompositions of a surface $\cal S$
that are subdivisions of a decomposition  of $\cal S$
into $4$-punctured spheres.   In terms of the analogy to
spin networks this corresponds to what has been called decomposing
a 4-valent node of a spin network in terms of two trivalent
nodes and an internal, or ``virtual", edge. 
In the present context all of these are connected by elements of
the modular group \cite{ms,louis2d3d}.  

Clearly, a given surface $\cal S$ has more than one
inequivalent decompositions into $4$-punctures spheres.
As an example, consider the tubular 4-simplex of Figure \ref{4simplex}.
The relationship between these different compositions correspond
to transformations between two bases in which 
the roles of the representations
and the intertwiners are exchanged.   This has interesting 
consequences for quantum geometry that we will discuss below,
when we describe how the geometrical interpretation of the theory
is constructed.

We will use the tubular evolution rules to define the dynamics
of the theory.  But first we have to define operators on
${\cal H}_{G_q}$ that implement them.

\section{Tube operators}

We now turn to the operators
on the space of states ${\cal H}_{G_q}$.  The
Moore-Seiberg \cite{ms} operators are a set of unitary operators 
that act inside
each ${\cal V}^{\cal S}$.  However, if our theory is to
be a generalization of spin networks there must be operators
that take us from states in one ${\cal V}^{\cal S}$
to states in another ${\cal V}^{{\cal S}^\prime}$ on 
a different surface $\cal S'$.  We will see here that several
useful sets of operators can be constructed, which will play a role
in the interpretation and dynamics of the theory. 
They are  analogous to the loop operators whose
algebra defines the loop representation of general relativity
\cite{lp1}.  Here, because the states are defined without any
reference to a background manifold, the operators are defined
relationally, in terms of decompositions of the surfaces
$\cal S$ into pieces.

Let $\Upsilon$ denote a genus $g$ compact oriented 2-surface with
$n \geq 1$ punctures $j_k$ ($k=1,..,n$). 
Given a compact $\cal S$  let 
$  r_I $ denote the maps 
\f
r_I :{\Upsilon} \rightarrow {\cal S}
\ff
taking ${\Upsilon}$ homomorphically to a  
component of ${\cal S}$. In general there will be a set of
such maps; they are distinguished by the index $I$.

   For each $I$ the map picks out a set of
$n$ non-intersecting circles $c_k^I$, $k=1,...,n$ in
$\cal S$.  Cutting $\cal S$ on these circles decomposes it into
the two pieces $r_I ({\Upsilon})$ and $\left({\cal S}- 
r_I ({\Upsilon})\right)$. The 
space of intertwiners ${\cal V}^{\cal S}$ decomposes as   
\f
{\cal V}^{\cal S}=\sum_{k}
{\cal V}^\Upsilon_{j_1...j_n} \otimes
{\cal V}^{\left({\cal S}-r_I ({\Upsilon})\right)}_{j_1...j_n}.
\ff
A state $|{\cal S}, \Psi \rangle \in {\cal V}^{\cal S}$
then decomposes to a sum over the representations
$j_1,...,j_n$ of the product of a state in 
${\cal V}^\Upsilon_{j_k}$ 
and a state in ${\cal V}^{\left({\cal S}-r_I ({\Upsilon})\right)}_{j_k}$,
\f
|{\cal S}, \Psi \rangle 
 =\sum_{k }|{\Upsilon}, {j_k} , \Psi^1 \rangle 
 \otimes
|\left({\cal S}-r_I ({\Upsilon})\right) , j_k , \Psi^2 \rangle.
\label{decom2}
\ff

Using this decomposition we then define three classes of operators.
The first two  
are block diagonal in the decomposition (\ref{states}), while
the third changes
the topology of the surface $\cal S$.

\subsection{Surface operators}

Let $F(\{j_k\} )$ be a symmetric function of $n$ representation
labels $\{j_k\}$.  Given such a function and a 2-surface $\Upsilon$
with $n$ punctures, 
 there is an hermitian 
operator $\widehat{\cal F}_\Upsilon$ that acts in
${\cal H}_{G_q}$ as follows.   
On the spaces 
${\cal V}^{\Upsilon}_{j_1...j_n }$ , $\widehat{\cal F}_{\Upsilon}$
is  the diagonal operator equal to $F(\{j_k\} )$.  Then
on a general state
\f
\widehat{\cal F}_\Upsilon|{\cal S}, \Psi \rangle 
 =\sum_I 
\sum_{k }F(\{j_k\} ) |{\Upsilon}, {j_k} , \Psi^1 \rangle 
 \otimes
|\left({\cal S}-r_I ({\Upsilon})\right) , j_k , \Psi^2 \rangle.
\label{decom3}
\label{surface}
\ff

This operator looks for the instances of the submanifold
$\Upsilon$ in each surface $\cal S$ and, in each state, measures
a property of the boundary separating $\Upsilon$ from the
remainder ${\cal S}-r_I(\Upsilon)$ given by the function 
$F(\{j_\alpha\} )$.  
The punctured surface
$\Upsilon$ can be thought of as 
the algebraic representation of a 3-dimensional region
$\cal R$ in the quantum geometry that  a state in 
${\cal V}^{\cal S}$ represents.
The surface
operators thus measure properties of  boundaries of regions
in space.
This interpretation will be developed in the next section,
where we will see that an 
example in the case of $SU(2)$ is given by the area  
operator obtained in quantum general relativity 
\cite{spain,volume1}.  

\subsection{Bulk operators}

Once a region $\cal R$ of an abstract quantum geometry has
been identified by a map $r_I: {\Upsilon}\rightarrow {\cal S}$
we can also try to measure bulk properties of that region.
These will be eigenvalues of operators that act on the 
space ${\cal V}_{j_1...j_n}^\Upsilon$ and depend  on the 
topology of $\Upsilon$ and 
hence on ${\rm dim}{\cal V}^\Upsilon_{j_1...j_n}$.  
To define such an operator let
us choose, for every ${j_1...j_n}$ an operator 
${\widehat B}_{j_1...j_n}$  
on ${\cal V}^{\Upsilon}_{j_1...j_n}$.  The
corresponding bulk operator  $\widehat{{\cal B}}^{\Upsilon}_{j_1...j_n}$ 
is defined on the state space
${\cal H}_{G_q}$  as
\f
\widehat{{\cal B}}^{\Upsilon}_{j_1...j_n}
|{\cal S} , \Psi \rangle
= \sum_I \sum_{k}
{\widehat B}
|{\Upsilon}, j_k,  \Psi^1 \rangle  
 \otimes
|\left({\cal S}-r_I({\Upsilon})\right) , j_k, 
\Psi^2 \rangle.
\label{bulk}
\ff

Examples of bulk operators are the volume operators which we will
describe in the next section.

\subsection{Substitution operators}
\label{subst}

In the last section we defined the tubular evolution moves.
These are examples of a large class of substitution operations
that take us from one manifold $\cal S$ to a different manifold
${\cal S}^\prime$ by cutting out a piece,
$\Upsilon^1$ of $\cal S$   and replacing it with a
different manifold $\Upsilon^2$ with the same boundary.  
The tubular evolution moves are examples of these.  
For such substitutions we 
can define linear operators that act on ${\cal H}_{G_q}$  
and take states from ${\cal V}^{\cal S}$ to those in
${\cal V}^{{\cal S}^\prime}$. 

Start with two punctured surfaces, ${\Upsilon}_1$ and ${\Upsilon}_2$,
each with an ordered set of $n$ punctures, with labels
$j_1,...,j_n$. They can be represented by vector spaces 
${\cal V}^{\Upsilon_1}_{j_1...j_n}$ and
${\cal V}^{\Upsilon_2}_{j_1...j_n}$.  Note that in
general  
${\rm dim}{\cal V}^{\Upsilon_1}_{j_1...j_n} \neq
{\rm dim}{\cal V}^{\Upsilon_2}_{j_1...j_n}$.
Given two vector spaces, we have the space of linear maps
from the first to the second, denoted
${\rm hom}\left({\cal V}^{\Upsilon_1},
{\cal V}^{\Upsilon_2}\right)$.  A particular
$\widehat c\in{\rm hom}\left({\cal V}^{\Upsilon_1},
{\cal V}^{\Upsilon_2}\right)$  acts on a state,
$|\Upsilon_1,j_k,\Psi^1\rangle
\in {\cal V}^{\Upsilon_1}_{j_1...j_n}$ as 
\f
\widehat c |\Upsilon_1,j_k,\Psi^1\rangle=
|\Upsilon_2,j_k,\Psi^2\rangle\in 
{\cal V}^{\Upsilon_2}_{j_1...j_n},
\ff
giving a state in ${\cal V}^{\Upsilon_2}_{j_1...j_n}$.
Given any such $\widehat c$ 
we construct  a substitution operator
$\widehat{\cal C}_{\Upsilon_1,\Upsilon_2,\hat c}$, defined by
\f
\widehat{\cal C}_{\Upsilon_1,\Upsilon_2,\widehat c}|{\cal S},\Psi\rangle
=\sum_I\sum_k|\left({\cal S}-r_I(\Upsilon_1)\right),j_k,\Psi^1\rangle
\otimes\left[
\widehat c|\Upsilon^1,j_k, \Psi^1\rangle\right].
\ff  
The action of 
$\widehat{\cal C}_{\Upsilon_1,\Upsilon_2,\widehat c}$ 
is pictured in Fig.\ \ref{move}.

Note that we may also glue $\Upsilon_1$ and 
$\Upsilon_2$ along their identical boundaries as in the right 
hand figure of Fig.\ \ref{move2}.

\section{Geometrical interpretations}

So far, we have defined states in ${\cal H}_{G_q}$ in terms
of labelled 2-dimensional manifolds.  We shall now 
interpret them in terms of observables 
related to  3-dimensional space.  These arise as natural
extensions of the observables of quantum general relativity:
the area and volume operators.  

The subtlety is that here 
there
is no background manifold. All of the properties of space, including its
topological and metric properties, must be coded into the states.  In
the absence of any background manifold to provide surfaces
and regions,  geometrical observables are constructed 
relationally,
from information coded into the states.  

Let us begin with the space of states ${\cal V}^{\cal S}$
associated to a given 2-surface $\cal S$. A 
{\it microscopical geometrical interpretation} of these 
states exists for every  
decomposition of $\cal S$ into a set of  $n$-punctured 2-spheres,
$B_I^n$, with $n \geq 3$, 
joined on a set of circles, $c_{\alpha}$.  Let us consider a
basis of states which is (partially) determined by definite values
for the representations $j_{\alpha}$ for these circles.  This state
is of the form, 
$|{\cal S},j_\alpha , \mu_I\rangle $ 
with intertwiners $\mu_I \in {\cal V}^{B^n_I}_{j_1...j_n}$
for each of the  punctured
spheres.   

The geometrical interpretation is constructed as follows.  Associated
to each   $B^n_I$ 
is a region ${\cal R}_I$.  These regions have three kinds of properties:

\begin{itemize}

\item{}{\bf Surface properties:}  A surface property of a region
${\cal R}_I$ is a function $F(j_1,...,j_n)$ of the labels on
the punctures of the corresponding  
$B^n_I$. Surface properties are measured by surface operators (\ref{surface}).

\item{}{\bf Bulk properties:} A bulk property of a region ${\cal R}_I$
is measured a hermitian operator $\widehat{\cal B}$ (\ref{bulk}) 
in the space of intertwiners
${\cal V}^{B^n_I}_{j_1...j_n}$.

\item{}{\bf Shared properties:} Two regions ${\cal R}_I$ and 
${\cal R}_J$ may have shared properties if they have a set of
common punctures with labels, say, $j_1,...,j_k$.  If this set is
non-empty, then $j_1,...,j_k$ is the 
common boundary of ${\cal R}^I$ and 
${\cal R}^J$.  A shared property of ${\cal R}^I$ and 
${\cal R}^J$ is then a function $G(j_1,...,j_k)$.

\end{itemize}

In the $SU(2)$ case we may import the kinematical structure
from quantum general relativity found in 
\cite{spain,volume1,sn1} to give us examples of each kind
of observable:

\begin{itemize}

\item{}The area of  ${\cal R}_I$ is a surface property.
It is given by 
$F(j_1,...,j_n )= l_{Pl}^2 \sum_{\alpha=1}^n
\sqrt{j_\alpha (j_\alpha +1)} $.

\item{}The volume of the interior of $B^n_I$ is an example of a
bulk property.  As we know from 
\cite{volume1,othervolume} the volume operator
is a hermitian operator $\widehat V [j_1,...,j_n]$ that acts in
the space or intertwiners ${\cal V}^{B^n_I}_{j_1,...,j_n}$ 
\footnote{Here we take the
definition given in \cite{volume1} that does not 
require any assumptions
about structure not present in our case such as 
linear relations
among tangent vectors at the nodes.}

\item{}The area also gives an example of a shared property.
If $B^n_I$ and $B^m_J$ share a set of $k$ spins $j_1,...,j_k$ then the
area of the common boundary of ${\cal R}_I$ and ${\cal R}_J$
is given by $l_{Pl}^2 \sum_{\alpha=1}^k
\sqrt{j_\alpha (j_\alpha +1)} $ summed over the common punctures
of the two regions.

\end{itemize}
 
Note that, given a division of $\cal S$ into punctured spheres, we may 
simultaneously
diagonalize all of the area and volume operators on the
corresponding regions ${\cal R}_I$.  Thus, a common 
eigenstate
$|{\cal S},j_\alpha ,\mu_I \rangle $ may be called a {\it 
microscopic quantum geometry}.  It is a set of regions
together with i) an area for the boundary of each one,
ii) an area for each common boundary,  
such that the area of each is the sum of its
common boundaries with the others and iii) a volume for each 
region.

For a general $G_q$, we expect a {\it generalized
microscopic quantum geometry} to be the maximal set of 
simultaneous
eigenvalues of surface and bulk observables for a 
decomposition
of $\cal S$ into punctured spheres.  There is also a notion of
a coarse-grained quantum geometry.  
We will discuss this in section 8.

\subsection{Duality between edges and intertwiners}

The reader may have noticed that the geometrical 
interpretations
available to the states in ${\cal V}^{\cal S}$ are not determined
by $\cal S$.  There is a geometrical interpretation for every
way of dividing $\cal S$ into punctured spheres.  We regard this 
freedom
as an intrinsic and attractive feature of the generalization from spin 
network
states to the space of states ${\cal H}_{G_q}$.  
For example, in the
4-valent spin networks 
two kinds of edges appear: real edges
connecting the nodes and ``virtual" edges that may be used to 
label
the intertwiners of the 4-valent nodes.  Thus, in the usual spin network
formalism they play different roles.  

\begin{figure}
\centerline{\mbox{\epsfig{file=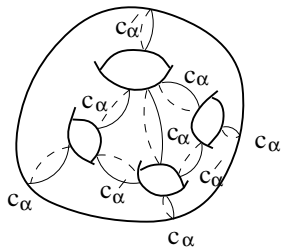}}}
\caption{ A choice of decomposition into 4-punctured spheres where
$c_\alpha$ are representations and $c_\alpha' $ intertwiners.}
\label{caca}
\end{figure}

For example, consider the tubular 4-simplex $\cal P$ and the
two different decompositions into 4-punctured spheres
illustrated in Figure \ref{4simplex}.   These may be described in terms
of two sets of circles $c_\alpha$ and $c_{\alpha^\prime}$,
as shown in Figure \ref{caca}.  (The full set 
$(c_\alpha, c_{\alpha^\prime})$ make up a maximal set of
non-intersecting circles on $\cal P$ and define a trinion 
decomposition of $\cal P$.)  
This decomposition represents states $|{\cal P},j(c_\alpha),
\mu_I)\rangle$. If we now read the decomposition with a different
set of circles, including
$c_\alpha'$, separating the five 4-punctured spheres, we obtain 
a different set 
of basis states $|{\cal P},j(c_\alpha^\prime),\mu_{I^\prime}\rangle$.
This shows that the
distinction between spins and intertwiners in this formalism 
is dependent on the choice of $n$-punctured spheres. Therefore,
so is the geometrical interpretation. 

\section{Causal evolution}

We now discuss the evolution of the
states in ${\cal H}_{G_q}$. The dynamics of the theory
will be based on the evolution moves defined in section
4.2.  By composing the moves we produce
sequences of states that we call histories.
These discrete histories share three characteristics
of Lorentzian spacetimes. 
i) There is a set of events which is a discrete
partially ordered set with no closed causal loops. This is a 
discrete analogue
of a Lorentzian spacetime.
ii)  There are connected sets of causally unrelated events,
the combinatorial analogues of spacelike surfaces.  iii)
A history can be decomposed in many ways into
sequences of spacelike surfaces, leading to a discrete analogue of
many fingered time.

\subsection{The evolution operator}

The evolution of states is generated by an operator that implements
the evolution moves described in section 4.2.  This
will be a substitution operator of the form defined in section 5.3.
To do this let $\rho=1,2,3,4$ correspond to the four kinds
of Pachner moves 
$1\rightarrow 4,  2\rightarrow 3, 3\rightarrow 2, 4 \rightarrow 1$.  
Then take $L_\rho$ to be the elementary local region
consisting of $\rho$ 4-punctured spheres, so
that $L_\rho^\prime$, the complement of $L_\rho$ in
$\cal P$ consists of $5-\rho$ 4-punctured spheres.
We will call $L_\rho$ the past set and $L_\rho^\prime$
the future set of the $\cal P$ associated with the 
$\rho \rightarrow 5-\rho$ move (See Figure \ref{move3}.).  

For $\rho=1$ and $4$ the $L_\rho$ and $L_\rho^\prime$
each have $4$ punctures, which are labeled by
representations $j_\gamma$, $\gamma=1,...,4$.  For
$\rho=2,3$ there are six punctures and $\gamma=1,...,6$.
For each $\rho$ and sets of $4$ or $6$
representations $j_\gamma$ we may choose operators 
$\widehat{c}_{\rho , j_\gamma}\in 
{\rm hom}({\cal V}^{L_\rho}_{j_\gamma}, 
{\cal V}^{L_\rho^\prime}_{j_\gamma})$.
The $\rho$'th move is then implemented by the
substitution operator
\begin{eqnarray}
\widehat{H}^\rho |{\cal S},\Psi \rangle &=&
\widehat{\cal C}_{L_\rho,L_\rho^\prime ,\widehat{c}_\rho} 
|{\cal S},\Psi \rangle \nonumber \\ 
&=&
\sum_I\sum_{k=1}^{n_\rho}|\left({\cal S}-r_I(L_\rho)\right),
j_\gamma,\Psi\rangle\otimes\left[\widehat c_\rho|L_\rho,j_\gamma,\Psi\rangle
\right].
\label{rhomove}
\end{eqnarray}
The total evolution operator is then given by 
\f
\widehat{H} = \sum_\rho \widehat{H}^\rho
\ff 

To see how these act, 
let us start with an initial state $|{\cal S}, \Psi \rangle$  
and act on it with one of the $H^\rho$.  
If ${\cal S}$ is large enough, there will be numerous regions
in it homeomorphic to 
$L_\rho$.  To each of them there is a map
$r_I: L_\rho \rightarrow {\cal S}$.  
For each $I$ we then cut from $\cal S$ the region
$r_I (L_\rho)$ and replace it by $L_\rho^\prime$.  This 
results  each time in a new
2-surface which we call ${\cal S}_{\rho , I}$. 
The result of the application of $\widehat{H}^\rho$  
is then a 
superposition of the states given by the action 
(\ref{rhomove}).  The exact map from the old states to the
new states is given by the linear maps
$\widehat{c}_\rho$. (We suppress the dependence of
$\widehat{c}_\rho$ on the representations $j_\gamma$.)

The operator $H$ is hermitian when each of the
$\widehat{c}_\rho$ are appropriately chosen.  
In this case a formal unitary evolution operator may be
written down as
\f
\widehat{\cal U}= e^{\imath \widehat H t}
\ff
where $t$ is a parameter having nothing to do with
the physical time (it just scales the operators $\widehat{c}_\rho$.)
The amplitude for 
an initial state 
$|initial \rangle = |{\cal S}_{initial}, \Psi_{initial}\rangle $
to evolve to a final state 
$|final \rangle = |{\cal S}_{final}, \Psi_{final}\rangle $
is formally given by 
\f
{\cal A}[|initial\rangle \rightarrow |final \rangle ] 
= \langle final |\widehat{\cal U} | initial \rangle . 
\label{a1}
\ff

\subsection{Amplitudes for causal evolution
and a discrete path integral}

By decomposing the action of  $\widehat{\cal U}$ 
at each order $n$ of the action of $\left(\widehat{H} \right)^n$ in terms of 
4-punctured spheres produced by the evolution moves, the
amplitude (\ref{a1}) can be given in terms of a sum over
a set of histories,  ${\cal M} = \{ |1\rangle,
 |2\rangle,|3 \rangle .... \}$ in which each $|I+1 \rangle$ results
from the previous $|I\rangle$ by the application of
one of the four moves.  
The theory gives an amplitude to 
each transition from an initial 
state to one of its successor states.
The amplitude is given by 
\f
{\cal A}_{L \rightarrow L^\prime} =  
\langle L^\prime , j_{k^\prime} , \mu_I^\prime 
|\widehat{c}^\rho |L , j_k , \mu_I\rangle,
\label{amp}
\ff
where $|L , j_k , \mu_I^\rangle$ is a trinion basis state 
for the initial elementary
local region to be cut out and 
$|L^\prime , j_{k^\prime} , \mu_I^\prime   >$ is a basis 
state on the elementary local region that replaces it.

Consider now an 
$(N-1)$-step history ${\cal M} = \{|1\rangle, |2\rangle,...,|N\rangle\}$.
Each transition is a generalized evolution move which has
an amplitude ${\cal A}^I$ given by
(\ref{amp}) for the transition from $|I\rangle$ to $|I+1\rangle$, 
$I=\{1,...,N-1\}$.  The
amplitude of the history $\cal M$ is then given by
\f
{\cal A}[{\cal M}] =\prod_I {\cal A}^I.
\ff
Let us then have two states,
 $|initial\rangle$ and $|final\rangle$.  There is an
infinite number of histories $\cal M$ such that the first
state is equal to $|initial\rangle$ and the last state is equal to
$|final\rangle$.  By analogy to the simplical case we may denote this as
$\partial {\cal M} = |initial \rangle \cup |final \rangle$.
The transition amplitude to evolve to $|final\rangle$ given
$|initial\rangle$ is then,
\f
{\cal A}[|initial\rangle \rightarrow |final \rangle ] = 
\sum_{{\cal M}|_{ \partial {\cal M} = |initial \rangle \cup |final \rangle}}
{\cal A}[{\cal M}].
\ff

As this is an infinite sum one may first compute the amplitude
for $|initial\rangle$ to evolve to $|final \rangle$ in $N$ steps.  This is
given by
\f
{\cal A}^N [|initial\rangle \rightarrow |final \rangle ] = 
\sum_{{\cal M} |_{|1\rangle  = |initial \rangle, |N\rangle= |final \rangle}}
{\cal A}[{\cal M}],
\label{an}
\ff
i.e., the sum over $(N-1)$-step histories that take 
the initial to the final state.  However, note that while
the full amplitude (\ref{a1}) 
is formally unitary by construction the same is not the
case for the $N$ step amplitude (\ref{an}).

\subsection{The causal structure}

We now show that each history $\cal M$ has defined
on it a discrete causal
structure as a result of its construction from the evolution moves.
Each history consists of $N$ states $|I\rangle$ which are
elements of ${\cal H}_{G_q}$.  Furthermore, the states
$|I\rangle$ come as labeled spin-tubes. Each one has
a set of descriptions in terms of generalized areas and volumes
because of its decompositions into $n$-punctured spheres.
Each history may be thought of as consisting of a 
succession of quantum $3$-geometries.   
Besides the representations and intertwiners, there is another
structure defined on the histories:  
each history $\cal M$ is a causal set, whose structure
is determined as follows.  

Each history $\cal M$ is also a set of  genus-6 elementary
spin-tubes ${\cal P}_i$.  Each ${\cal P}_i$ is divided into two parts
$L_i$ and $L_i^\prime$ corresponding to the elementary 
local regions that were removed and inserted.
The 4-punctured spheres in $L_i$ are the 
{\it past set} of ${\cal P}_i$. The remaining
4-punctured spheres, which are in
the complement $L^\prime_i$ are the {\it future set} of 
${\cal P}_i$.  
Now, consider a particular $4$-punctured sphere $s$ in the future 
set $L_i'$ in some ${\cal P}_i$.  Let us assume that $s$ 
 has been acted on by at least
one generalized evolution move 
${\cal P}_j$ for $j > i$.  Then $s$ also belongs to the past subset 
$L_j$ of ${\cal P}_j$. 
If now $s'$ is a 4-punctured sphere in the future subset $L_j'$ of
${\cal P}_j$, we will say
that $s'$  is to the {\it immediate causal future} of $s$.  

Now, consider a sequence of $r$ 
4-punctured spheres $s_i$, $i=1,...,r$, such that for each
$s_i, i < r$ either i) $s_{i+1}$ is to the immediate causal
future of $s_i$, or ii) there is some 
$|I\rangle \equiv |{\cal S}_I , \Psi_I \rangle \in {\cal M}$
such that $s_i$ and $s_{i+1}$ are both in the surface
${\cal S}_I$ and $s_i \cap s_{i+1} \neq 0$.  
(This, is either each 
4-punctured sphere in the sequence is to the immediate
causal future of its predecessor, or it and its predecessor
overlap in a single surface associated with a state
$|I\rangle$ in the history.)
When this is the case we will say that $s_r$
is to the causal future of $s_1$, $s_r > s_1$.

It is clear that the relation $>$ is transitive and that given
two 4-punctured spheres $s_1$ and $s_2$,
$s_1 > s_2 > s_1$ is never the case.
Thus, the 4-punctured spheres 
in each history $\cal M$ constitute a causal set, which
is defined in \cite{sorkin} to be a partially ordered set 
with no closed causal loops which is locally finite.  The latter
means that given any $s_1$ and $s_2$ the set contained in
the causal past of $s_2$ and the future of $s_1$ is finite.
As argued in \cite{sorkin,thooft} a discrete set that has
on it a causal structure is a candidate for a discrete 
model of spacetime.

The 4-punctured spheres of a history $\cal M$, defined
by the evolution moves that construct it, are then the {\it events}
of ${\cal M}$. We will call the set of events ${\cal E}$. 
By construction, $\cal E$ is a causal set. It differs from the causal
set of Sorkin and collaborators\cite{sorkin} in that there is
additional structure, associated to a notion of space.

Each history $\cal M$ may be foliated by a number of sets
of causally unrelated events of $\cal M$ that we will call
the
spacelike slices $\Gamma$.  A spacelike slice of $\cal M$
is a subset $\{s_a\}$ of
$\cal E$ glued together according to the following rules:
\begin{enumerate}
\item{}
No two $s_a$ in $\Gamma$ may be causally related.
\item{}
Two events $s_a$ and $s_b$ in $\Gamma$ may be
glued together if there is a state $|I\rangle \in {\cal M}$ in which
they are glued along some circle.  
\item{}
The set $\Gamma$ is maximal in that no $s_a$ may be added
to it without violating these conditions.  
\end{enumerate}
Associated with $\Gamma$ is a state
$|\Gamma \rangle \in {\cal H}_{G_q}$  given by
$|{\cal S}_{\Gamma}, j, \mu_a  \rangle$.  Here the
intertwiners $\mu_a$ are fixed because 
$s_a \in \Gamma$ are given. 
Similarly, each circle $c_{ab}$ along which two
adjacent $4$-punctured spheres $s_a$ and $s_b$ are glued
is in fact a circle labeled by a fixed representation $j$.
Hence the labels on the state
$|\Gamma \rangle = |{\cal S}_{\Gamma}, j, \mu_a  \rangle$ are 
uniquely determined
by the history $\cal M$.

The $N$ original states $\{|1\rangle, |2\rangle,...,
|I\rangle,...,|N\rangle\}$  are spacelike slices according to this
definition.  But there are many more sequences which may be constructed
given the history ${\cal M}=\{|1\rangle, |2\rangle,...,
|I\rangle,...,|N\rangle\}$ that have $|1\rangle$ as the 
initial state and $N\rangle$ as final.  We call the 
set of such states ${ W}_{\cal M}$.  One may in general
select other sequences of elements of ${ W}_{\cal M}$, e.g.\ ${\cal M}'=
\{|1\rangle, |2^\prime \rangle , 
...,|I'\rangle,...,|N\rangle\}$, 
 that have the
property that every event in $\cal E$ is a $4$-punctured 
sphere in a decomposition of at least one $|I^\prime \rangle$.
As far as the local geometry and causal structure are concerned
these are equivalent descriptions of the history
$\cal M$.  Thus, this quantum theory has a discrete analogue
of multi-fingered time.

Thus, a discrete history $\cal M$ combines
discrete analogues of both 
the canonical picture of quantum gravity and the
spacetime causal structure.  It is the marriage
of both kinds of structure within a completely
discrete approach to quantum gravity that we believe
gives this approach its particular power.

\subsection{Connection with spin foam and membranes}

In a number of recent papers, 
\cite{mike,carlomike,louis,louisjohn,baez} a concept Baez calls
``spinfoam" has been introduced.  These are networks 
of colored 
2-surfaces embedded
in a four-dimensional spacetime whose slices by 
three-manifolds are spin networks.  Gupta\cite{sameer} has
shown that the spin foam can be given a Lorentzian
formulation by the addition of a causal structure 
and that that formulations is in a 
particular sense dual to the formulation of \cite{fotini1}. 
There is an analogous spacetime foam
structure associated with the histories $\cal M$, although it
has not been so far investigated.  It can be constructed by noting
that each of the evolution moves may be seen as three-dimensional
cobordisms between the two surfaces $L_\rho$ and 
$L_\rho^\prime$ (See Fig.  \ref{move2}). 
The resulting three-manifolds  
may be joined together  to
construct a three-dimensional timelike combinatorial
manifold associated to each history $\cal M$.   This is
a non-perturbative,
background independent membrane.

\section{Coarse graining, entropy and the holographic
hypothesis}

Before closing we make some comments about 
coarse graining and entropy that will enable us to comment
also on
the relationship
of our proposal to the holographic hypothesis\cite{holo-thooft,holo-lenny}
and the Bekenstein bound\cite{bek}.   

The basic idea is that in addition to the fine grained observables
discussed previously there are coarse grained observables
that describe statistical information about the
states defined in section 2. 
There are two kinds of course grainings which are
relevant.  In the first we retain information about the
topology of the surface $\cal S$ while in the second 
we retain only information that can be measured
by observers at the boundaries of the regions.

Before describing these we may note that
the existence of coarse grained observables in itself means that
the theory genuinely has local observables that are not determined
by the values of the coarse grained observables.

\subsection{Coarse graining by topology}

We can coarse grain the information in a state
$|{\cal S}, \Psi\rangle$ by forgetting the information
about the state $\Psi \in {\cal V}^{\cal S}$ and retaining
only statistical information about the surface $\cal S$.
This results in a density matrix which is constructed by
tracing over the representations $j_\alpha$ and
intertwiners $\mu^\rho_I$. 
To each surface $\cal S$ is then associated a density matrix
which is $\rho_{\cal S}=P_{\cal S} $, the projection
operator onto
${\cal V}^{\cal S}$.  
There is an entropy associated with this coarse graining.
Associated to each surface $\cal S$ is an entropy
$S[{\cal S}]= ln({\rm dim}{\cal V}^{\cal S} )$.  

As the dynamics changes the topology an entropy change
can be associated with the evolution operators defined in
the last section.  This makes possible a thermodynamic treatment
of the evolution, which will be described elsewhere.

\subsection{Coarse graining by regions}

Rather than coarse graining by the topology of $\cal S$ we
can coarse grain by splitting space into regions and measuring
statistical information about each region.  To do this we must
take into account what we learned from our discussion of
geometrical interpretations, which is 
that as the topology and geometry of space are defined from
the states, the splitting of space into regions must be defined
intrinsically in terms of the states.  We then  
define a coarse grained quantum geometry as
a coarse grained interpretation of a quantum
state $|{\cal S}, \{j\} ,\{\mu \} \rangle$.  Let us then
consider a decomposition of $\cal S$ into a set of 
regions
$R_i$ along $m_i$ circles $c_\gamma$.  Each
piece consists of a component of $\cal S$ we will
call $W_i$.  Each $W_i$ is a punctured surface,  punctured
by the $m_i$ labels $j_\gamma$ on the circles $c_\gamma$.

To each region we will also associate a punctured $S^2$,
with $m_i$ punctures with the same labels as the $S_i$.  
Coarse graining will mean that for each region $R_i$ we forget
the details of the topology of the component $W_i$. This means
that all observables concerning the region must be representable
as operators in the space of intertwiners on the associated 
punctured $S^2$.  
There are then two spaces of intertwiners which are relevant,
${\cal V}^{W_i}_{j_\gamma}$ and ${\cal V}^{S^2}_{j_\gamma}$.
Coarse graining consists of replacing a microscopic state,
which is a vector in ${\cal V}^{W_i}_{j_\gamma}$
with a density matrix in ${\cal V}^{S^2}_{j_\gamma}$.
 
In correspondence with the
different notions of properties we may define a {\it coarse
grained surface property} of the region  
$W_i$ to be a function of the labels $j_1,...,j_{i_m}$ and a
{\it coarse grained bulk property} to be an operator
in ${\cal V}^{S^2}_{j_\gamma}$.  Finally, two regions may
share properties when the corresponding $W_i$'s
are glued along punctures.  Moreover, given a full set of labelings
on the punctured surface $W_i$ we have a state in
${\cal V}^{S^2}_{j_\gamma}$ by considering the $W^i$ as a framed spin network
embedded in the interior of the surface $S_i$ in $R^3$.

A coarse grained description of the quantum geometry is
then given by a density matrix in the spaces 
${\cal V}^{S^2}_{j_\gamma}$ that corresponds to each of the
regions $R_i$.  It corresponds to what observers may measure
about the world, assuming they can only measure on boundaries.

\subsection{Connection with the holographic hypothesis and
Bekenstein bound}

The possibility of describing coarse grained properties
in this way  also
suggests a formulation of the holographic\cite{holo-lenny,holo-thooft} 
hypothesis that is
entirely non-perturbative and background independent. 
This arises in the case that we split the universe into two regions,
and assume that we can only make measurments in one of them.

Let us introduce a splitting of a surface $\cal S$ along a set
of $p$ non-intersecting elements of $\pi^1 [{\cal S}]$,
which we will call the  $c_\gamma$, $\gamma = 1,...,p$.   
The two halves may be
called ${\cal S}^+$ and ${\cal S}^-$; the
$c_\alpha$ are in each case their ends.  Let us further consider
a basis of states in which there are definite representations
$j_\gamma$ defined on the surfaces.  

In the absence of a background manifold we will 
simply represent  the splitting by a $p$-punctured $S^2 $, 
labeled by the $j_\gamma$.  Each half ${\cal S}^\pm$ 
then has on it a space of intertwiners 
${\cal V}^{{\cal S}^\pm}_{j_\gamma}$.
An element   ${\cal V}^{{\cal S}^\pm}_{j_\gamma}$ defines 
what we will call a {\it quantum
geometry with boundary}.    Given
a quantum geometry, i.e.\ a state in a  ${\cal V}_{\cal S}$,
there are many ways to split it into two halves,
giving two quantum geometries with boundaries.  
The splitting of the world into two parts constitutes a simple 
coarse graining of it.

Now consider an observer who lives in one  
a half, ${\cal S}^+$, who is for some reason unable to measure 
any information about the
topology or state of $\cal S$ in the other half ${\cal S}^-$.  
This might, for example,  arise if the
causal structure (which we have shown makes 
sense at this, non-perturbative background
independent level) does not enable him or her to receive
any information from the other half.  In this case the observer
effectively lives in a quantum geometry with boundary defined
by the half ${\cal V}^{{\cal S}^+}_{j_\gamma}$.  

What information can the observer have about the physics
of the other half ${\cal V}^{{\cal S}^-}_{j_\gamma}$?  All they
can measure is correlations between measurements they may
make at the $p$ ends.  This means that the possible states
they may distinguish by their measurements are given exactly
by the space of conformal blocks on the $p$-punctured
$S^2$ associated with their boundary.  This is the space
${\cal V}^{S^2}_{j_\gamma}$ which we described before.  

To summarize,  the following may be considered 
a {\it non-perturbative formulation
of the holographic hypothesis}:   
When
an observer is unable to measure information corresponding to
the interior of a region of a quantum geometry, because of the
presence of a causal horizon, or for any other reason, the information
accessible to them by measuring observables at the boundary of
that region is represented by a finite dimensional space of
states ${\cal V}^{S^2}_{j_\gamma}$ for
some p-punctured $S^2$.  

This has several further consequences.  First, in the
$SU(2)$ case  it is known that\cite{linking}
\f
\ln \left ( {\rm dim} [ {\cal V}^{S^2}_{j_\gamma} ]
\right ) \leq {c\over 4} {A[ j_\gamma ] \over l_{Planck}^2 }
\ff
for large numbers of punctures, where 
$c=8ln(2)/\sqrt{3}$.  Here 
$A[j_\gamma ]$ is the area operator of quantum general
relativity \cite{spain,volume1} with eigenvalues
$\sum_{\gamma} l_{Planck}^2 \sqrt{j_\gamma (j_\gamma+1)}$.  Thus,
{\it the Bekenstein bound\cite{bek} is automatically satisfied}
\footnote{We may note that the constant $c$ is not equal to one.  This
is not surprising given that the quantity $l_{Planck}$
in the area formula is given by the bare Newton's
constant.  Unless the theory has a continuum limit the macroscopic,
renormalized Newton's constant which plays a role in black hole
thermodynamics cannot be defined.  This result suggests then
predicts that in those cases $G_{ren}= c G_{bare}$.}.

In the case of a general $G_q$ we do not know
which observable corresponds to the area.  
It may be any surface property, which means it
must be an additive function $A$ of the casimers of $G_q$.  
The Bekenstein bound gives us a constraint
on that definition, which is that   
\f
A[ j_\gamma ] < 4 l_{Planck}^2  \ln \left ( {\rm dim} 
[ {\cal V}^{S^2}_{j_\gamma} ]
\right ). 
\label{genbek}
\ff

We may note that the Bekenstein bound (\ref{genbek}), together
with certain other assumptions is, as Jacobson has shown 
 \cite{tedthermo},  equivalent to the Einstein equations.  
Jacobson's argument in \cite{tedthermo} can be interpreted
to imply that 
any finite theory of quantum gravity 
that has a classical limit such that a) the relationship 
(\ref{genbek}) is satisfied on every horizon which exists by
virtue of an observer being accelerated and b) quantum
fields behave as conventional free fields in the limit of low
curvatures, then the field equations of general relativity are true
to leading order in curvatures as a consequence 
of the ordinary laws of thermodyanics\cite{tedthermo}.
This suggests that statistical assumptions about the dynamics,
together with (\ref{genbek}) may be sufficient to derive the
classical limit of the theory.

\section{Conclusion}
 
The general framework introduced here becomes a theory with
two inputs: a group or algebra $G_q$, and a choice of the 
dynamical operators $\widehat H^\rho$ that define the evolution.
The main question that must be investigated is how these 
operators are to be chosen. Good choices should 
lead to a theory with a good 
continuum limit which reproduces classical general relativity with
matter fields. This is currently being
investigated in several directions. 
 
\begin{enumerate}

\item{}The algebra of the tube operators introduced here should
be worked out. It will be interesting to see if there is a set
of local operators that generate the algebra and if they are
related to the loop algebra of quantum gravity \cite{lp1} in
the $q \rightarrow 1$ limit.

\item{}It appears possible to choose the evolution operator
$\widehat{c}^\rho$ to agree with the dynamics generated
by the lorentzian hamiltonian constraint of 
Thiemann\cite{qsdi}.  A path integral representation of Thiemann's lorentzian 
constraint, along the lines of \cite{carlomike}, may then be possible.

It seems that the evolution generated by
Thiemann's constraints is ultralocal\cite{trouble1,trouble2}.
However we may note that the evolution generated by the 
$1\rightarrow 4$ and $4\rightarrow 1$ moves are ultralocal  
in the sense that they do not lead to long-range propagation.
As suggested already in the Euclidean context in \cite{carlomike}
it follows that the other moves are necessary in order to
have long-range propagation.  
   
\item{}More generally, the relation of the causal theory to the euclidean
path integral approaches \cite{mike,carlomike,louis,louisjohn,baez} 
should be investigated. In this 
direction, Gupta in \cite{sameer} has formulated a causal spin foam. 
 
\item{}All of the above involve so far only the $SU(2)$ spin networks. 
The extension to 
other groups is important. The $SO(8)$ case is of special interest
because of its connection to supersymmetry and triality. It is currently
under investigation with Asok.  The general case of a supergroup
should be investigated.  

\item{}Two connections with string theory have been 
investigated. In \cite{pqtubes},
we take $G_q$ to be the projective group of the circle. Its 
representations are parametrised by relatively prime pairs of integers
$(p,q)$. The states in this case turn out to be combinatorial 
$(p,q)$ string networks \cite{sennet} whose dynamics is a simple case
of section 7. Second, in \cite{stringsfrom} perturbations of the $SU(2)$ theory
have been studied which are given by a (1+1)-dimensional system with
couplings determined by $\widehat{c}^\rho$. When the full theory has
a good continuum limit, the action for the $1+1$ system is given to
leading order by the Nambu action of bosonic string theory.  An argument
may be given that if, for some choice of $G_q$ and $\widehat{c}^\rho$
the induced $1+1$ dimensional theory is a consistent perturbative
string theory then the continuum limit of the non-perturbative theory
exists.

\item{}In  \cite{fmls1} we argued that the existence of a continuum limit
can be seen as a critical phenomenon which is analogous to 
directed percolation.  To investigate this we have invented a set
of simple models that have dynamical causal structure of the 
type described here\cite{models}.  Further, these models are 
discrete dynamical systems since
evolution proceeds by discrete local steps.  
This leads to proposals for the evaluation of the path
integrals proposed here which are discussed in \cite{stulee}.
Finally, given the remarks in the previous section, 
one may use statistical mechanics to make general
statements about the evolution of the states based
on the entropy $S[{\cal S}]$.  
 
\end{enumerate}

\section*{ACKNOWLEDGEMENTS}

We are grateful to Chris Isham for suggestions on constructing
the space of states of the theory and how to accommodate the
action of the evolution operators. 
While writing this paper we consulted papers of Louis Crane
on conformal field theory\cite{louis2d3d} and were 
struck by the resonance of
the kinematical framework of section 2 with ideas sketched there.
We would like to thank him and John Baez,  Sameer Gupta, 
and Carlo Rovelli for comments on a draft of this paper.
We also thank them and Stuart Kauffman and Mike Reisenberger 
for conversations
and encouragement. 
This work was supported by
NSF grant PHY-9514240 to The Pennsylvania State
University and a NASA grant to the Santa Fe Institute.

\end{document}